# A Survey on Various Data Hiding Techniques and their Comparative Analysis

**Harshavardhan Kayarkar*** *Corresponding Author*
M.G.M's College of Engineering and Technology, Navi Mumbai, India
Email: hjkayarkar@gmail.com
**Sugata Sanyal**
School of Technology and Computer Science, Tata Institute of Fundamental Research, Mumbai, India
Email: sanyals@gmail.com

**ABSTRACT:** With the explosive growth of internet and the fast communication techniques in recent years the security and the confidentiality of the sensitive data has become of prime and supreme importance and concern. To protect this data from unauthorized access and tampering various methods for data hiding like cryptography, hashing, authentication have been developed and are in practice today. In this paper we will be discussing one such data hiding technique called Steganography. Steganography is the process of concealing sensitive information in any media to transfer it securely over the underlying unreliable and unsecured communication network. Our paper presents a survey on various data hiding techniques in steganography that are in practice today along with the comparative analysis of these techniques.

## 1. INTRODUCTION

Internet came into existence in the late 1960s and 1970s out of the need to exchange research data among the researchers across different universities and also to enable communication in the battlefield to convey vital information which could prove advantageous in the war situations. Since the inception of the internet, the security and the confidentiality of the sensitive information have been of utmost importance and top priority.

The reason for this security and confidentiality is because the underlying communication network over which the transfer of sensitive information is carried out is unreliable and unsecured. Anybody with the proper knowledge and right applications can eavesdrop and learn of the communication and intercept the data transfer which could be very dangerous and even life threatening in some situations.

Ideally the internet and the communication network and the routing protocols should exhibit the following the properties:

- **Security:** Security is an important property of the internet. The internet should provide and preserve the confidential and sensitive information that flows through it. The security should be such that only the intended recipient of the information should gain access to it.
- **Distributed Operation**: The internet should be distributed rather than only residing on some centralized server. In the event of the crash the internet should not lose its functionality and continue performing efficiently.
- **Reliability**: Reliable communication is one of the vital properties of the internet. The internet should guarantee the reliable delivery of the information to the intended recipient.

- **Fault-Tolerance**: Fault-tolerance means the ability of the system to operate normally even in the events of failure. Internet should exhibit fault-tolerance so that it keeps on functioning even when there is failure in some part of the internet.
- **Quality of Service Support**: Quality of Service (QoS) is one of the crucial properties in terms of communication. Inter should provide QoS support to various applications and sensitive data and should prioritize them depending on the nature of the data.
- **Robustness**: Internet should be robust in the sense that it should continue functioning normally even in the presence of errors and unexpected situations like invalid input.

All the above mentioned properties are ideal and cannot be practically implemented in the structure and functioning of the internet as it comprises of many networks, different infrastructures: wired, wireless, ad hoc and various mobility models [1] [2] [3] [4] [5]. One such property that cannot be guaranteed in the internet is Security.

Due to the inability to guarantee security, various vulnerabilities exist in the network that can be exploited and gives rise to several security attacks. Some of the common security attacks are listed below.

- Impersonation or Spoofing: The main goal of this attack is to assume the identity of the person and convince the sender that it is communicating with the intended recipient.
- Man in the Middle attack: In this attack, the attacker makes independent connections with the two parties across the network making them believe that they are communicating privately, when in fact the communication is controlled and intercepted by the attacker.
- Traffic Analysis: In this process the attacker listens to the chatter on the communication network between two parties without interacting between them and tries to learn the information that they are sharing.

To mitigate these security vulnerabilities and facilitate seamless and safe transfer of data over the communication channel, techniques like cryptography, hashing, authentication, authorization, steganography are developed.

Our paper illustrates various data hiding techniques in steganography to enable the safe transfer of critical data over the unsecure network.

Steganography is sometimes erroneously confused with cryptography, but there are some notable and distinctive differences between the two. In some situations steganography is often preferred to cryptography because in cryptography the cipher text is a scrambled output of the plaintext and the attacker can guess that encryption has been performed and hence can employ decryption techniques to acquire the hidden data. Also, cryptography techniques often require high computing power to perform encryption which may pose a serious hindrance for small devices that lack enough computing resources to implement encryption.

On the contrary, steganography is the process of masking the sensitive data in any cover media like still images, audio, video over the internet. This way the attacker does not realize that the data is being transmitted since it is hidden to the naked eye and impossible to distinguish from the original media.

Steganography involves 4 steps:

a. Selection of the cover media in which the data will be hidden.
b. The secret message or information that is needed to be masked in the cover image.
c. A function that will be used to hide the data in the cover media and its inverse to retrieve the hidden data.
d. An optional key or the password to authenticate or to hide and unhide the data.

In this paper we present a survey on various data hiding techniques in steganography along with their comparative analysis.

The rest of the paper is organized as follows: Section 2 presents the survey on various data hiding techniques and related work. Section 3 performs the Comparative Analysis of the techniques discussed in Section 2. Finally Section 4 draws the Conclusion of the paper.

## 2. VARIOUS DATA HIDING TECHNIQUES

In this section we will be presenting the survey on various data hiding techniques in steganography to facilitate secure data transmission over the underlying communication network.

### 2.1. Data Hiding Techniques in Still Images

Nosrati et al. [6] introduced a method that embeds the secret message in RGB 24 bit color image. This is achieved by applying the concept of the linked list data structures to link the secret messages in the images. First, the secret message that is to be transmitted is embedded in the LSB's of 24 bit RGB color space. Next, like the linked list where each node is placed randomly in the memory and every node points to every other node in list, the secret message bytes are embedded in the color image erratically and randomly and every message contains a link or a pointer to the address of the next message in the list. Also, a few bytes of the address of the first secret message are used as the stego-key to authenticate the message. Using this technique makes the retrieval and the detection of the secret message in the image difficult for the attacker.

Kuo et al. [7] [8] [9] [10] presented a reversible technique that is based on the block division to conceal the data in the image. In this approach the cover image is divided into several equal blocks and then the histogram is generated for each of these blocks. Maximum and minimum points are computed for these histograms so that the embedding space can be generated to hide the data at the same time increasing the embedding capacity of the image. A one bit change is used to record the change of the minimum points.

Das et al. have listed different techniques to hide data [11] [12]. The authors have mainly focused on how steganography can be used and combined with cryptography to hide sensitive data. In this approach they have explained and listed various methods like Plaintext Steganography, Still Imagery Steganography, Audio/Video Steganography and IP Datagram Steganography which can be used to hide data. The authors have also elucidated the Steganalysis process which is used to detect if steganography is used for data hiding.

Naseem et al. [13] presented an Optimized Bit Plane Splicing algorithm to hide the data in the images. This method incorporates a different approach than the traditional bit plane splicing technique. In this approach instead of just hiding the data pixel by pixel and plane by plane, the procedure involves hiding the data based on the intensity of the pixels. The intensity of the pixels in categorized into different ranges and depending on the intensity of the pixel, the number of bits are chosen that will be used to hide data in that particular plane. Also, the bits are hidden randomly in the plane instead of hiding them adjacent to each other and the planes are transmitted sporadically thus making it difficult to guess and intercept the transmitted data.

Fu et al. presented some novel methods for data hiding in halftone images [14], [15]. The proposed method enables to hide huge amounts of data even when the original multitone images are unavailable by forced pair-toggling. The resulting stego-images have high quality and virtually are indistinguishable from the original image.

Dey et al. [16] have proposed a novel approach to hide data in stego-images which is an improvement over the Fibonacci decomposition method. In this method the authors have exploited Prime Numbers to hide data in the images. The main agenda is to increase the number of bit planes of the image so that not only the LSB planes but even the higher bit planes can be used to hide to data. This is done by converting the original bit planes to some other binary number system using prime numbers as the weighted function so that the number of bits to represent each pixel increases which in turn can be used hide data in higher bit planes. The authors have also performed a comparison of the Fibonacci decomposition method with the traditional LSB data hiding technique showing that the former outperforms the latter method and comparing Fibonacci Decomposition method with the proposed method which outclasses the former method. Also, the proposed method generates the stego-image which is virtually indistinguishable from the original image.

### 2.2. Data Hiding Techniques in Audio Signals

Kekre et al. proposed two novel methods to transfer secret data over the network by hiding them in the audio signals, thus generating a stego-audio signal [17] [18]. In the first method the authors hide the secret data in the LSB of audio by considering the parity of the sample, i.e. instead of directly replacing the digitized sample of the audio with the secret message, first the parity of the sample is checked and then the secret data is embedded into the LSB. This way it becomes even more difficult for the intruders to guess the bit or the data that is being transmitted. In the second approach, XORing of the LSB's is performed. The LSB's are XORed and depending on the outcome of this operation and the secret data that is to be implanted, the LSB of the sample data is changed or left unchanged. A different approach is followed by Kondo. Kondo [19] proposed a data hiding algorithm to embed data in stereo audio signals. The algorithm uses polarity of reverberations which is added to the high frequency signals. In this method the high frequency signals are replaced by one middle channel and then the data is embedded. The polarity of reverberations that is added to each channel is performed to adjust the coherence between these channels. The detection of the embedded data is done by employing the correlation between the sum and difference of the stereo signal. Also, original signal is not required to extract the hidden data by using this algorithm.

### 2.3. Data Hiding Techniques in IPv4 Header

To securely transmit the data over the network the Vasudevan et al. [20] used the analogy of the jigsaw puzzle. They insinuate to fragment the data into variable sizes instead of fixed size like the jigsaw puzzle and append each fragment of data with a pre-shared message authentication code (MAC) and a sequence number so that the receiver can authenticate and combine the received fragments into a single message. At the sender side every data fragment is prefixed and suffixed with a binary '1' and then XOR'ed with a Random number called the one-time pad and transmitted over the network. When the receiver receives the message it performs the exact opposite process of that to the sender and retrieves the intended message.

Ahsan and Kundur presented two novel approaches that exploit the redundancy in the IPv4 header of the TCP/IP protocol suite to convey the secret message over the communication channel without detection [21], [22], [23], [24]. In the first method, the FLAGS field containing the fragmentation information is used to conceal the data and transmit over the network. In the second technique 16-bit identification field of the header through chaotic mixing and the generation of sequence numbers is used to hide the data and convey the information to the recipient.

### 2.4. Data Hiding Techniques in Video Sequences

Li et al. [25] [26] suggested a data hiding technique based on the video sequences. This method implements an adaptive embedding algorithm to select the embed point where the sensitive data is to be concealed. The scheme functions by adopting 4x4 DCT residual blocks and determining a predefined threshold. The blocks are scanned in an inverse zigzag fashion until the first non-zero coefficient is encountered. The value of this coefficient is compared with predefined threshold and if it is greater than the threshold then that pixel is chosen to embed the data.

### 2.5. Data Hiding Techniques using DNA Sequences

Abbasy et al. [27] [28] introduced a scheme to enable secure sharing of resource in cloud computing environments. The proposed method employs DNA sequences to hide data. The process consists of two steps. In the first step a DNA sequence is selected and the binary data is converted into this DNA sequence by applying the pairing rules. This step, apart from converting the data also increases the complexity by applying the complementary rules and then indexing the garbled sequence. The second step involves the extraction of the hidden data from the DNA sequence where in exactly a reverse operation is performed to the first step.

### 3. COMPARATIVE ANALYSIS OF VARIOUS DATA HIDING TECHNIQUES

In this section a comparative analysis of different data hiding schemes in steganography is presented.

The authors in [6] [7] [13] [14] [16] have all presented techniques to hide data in the still images and generated stego-images as the output. In [6] the authors embed the data in RGB 24 bit color image by using the linked data structures where in, the data hidden in the image is linked with other data. The advantage of this method is that hiding the data randomly than sequential will make it difficult for the attacker to locate it and also without the authentication key the attacker will not be able to access the next piece of data in the image. Instead of using the whole image as the cover image, the authors in [7] have proposed a method that segments the image into blocks of equal sizes. Also, the process involved in this method is reversible hence there is no loss of hidden data. The approach followed in this scheme to conceal data is quite different. In this technique the histograms of the blocks of images is taken and they are shifted to minimum point of the histogram and then the data is hidden between these points. The improvement of this technique is that it provides higher capacity to hide data than the previous method.

In [13] optimized bit plane splicing method is implemented. In this method the intensity value of the pixel is divided into different planes and rather than using the traditional method of hiding the data into LSB of the pixel and plane by plane, the data in this approach is hidden based on the intensity of the pixels. The pixels are grouped based on the intensity and then number of pixels used to represent the data is chosen depending on the intensities. Also, rather than hiding the data sequentially in the planes, the data is hidden randomly and during the transmission of the data the planes are transferred randomly to make it difficult to intercept the data. The advantage of this technique is that by grouping the pixels according to the intensity more number of bits is available to represent the hidden data than just the LSB of the pixel. In [16] to increase and utilize the higher bit planes to hide the data a different approach from the one discussed earlier is employed. This is achieved by converting the original bit planes into some other binary number system using the prime numbers as the weighted function. This enables to use more number of bits to represent the hidden data.

The authors in [17] [19] use audio signals as the cover media to hide the sensitive data. In [17] the authors present two techniques to hide data in the audio signals. In the first method before hiding the data in the LSB of the sample of the audio signal the parity of the sample checked. This method makes the attacker difficult to guess the transmitted data. In the second approach, the LSB's are XOR'ed and depending on the result of this operation and the hidden data the LSB of the sample data is decided to be changed or left unchanged. In [19] a separate approach is followed where in the stereo audio signals are used to embed the data. In the proposed algorithm the polarity of reverberations is applied to the high frequency signals which are then replaced by one middle channel to embed the critical data.

Jigsaw-based approach [20] is used to transfer data over the communication channel securely. In this scheme the data is fragmented in block of variable sizes and a message authentication code (MAC) is used to authenticate each and every piece of data. Also, every message is prefixed and suffixed with a binary 1 along with XOR-ing the data with the randomly generated one-time pad. By fragmenting the data the attacker is unable to make sense of the data at the same time he cannot access the data unless he possess the authentication code for the data. A diverse approach is followed by the authors in [21]. In this scheme the redundant fields in the IPv4 header is exploited to mask the data. The fragment bit of FLAGS and the 16-bit identification fields are utilized to pass the delicate data over the communication network.

TABLE I below provides a brief summary of the data hiding techniques in steganography with their advantages.

## 4. CONCLUSION

In this paper we discussed about steganography and presented some notable differences between steganography and cryptography. We also surveyed various data hiding techniques in steganography and provided a comparative analysis of these techniques.

In the Introduction section we discussed about various security flaws and vulnerabilities in internet. We also discussed about various techniques to enable the secure transfer of data with the help of methods like cryptography, steganography, hashing, and authentication. In the next section we presented various techniques to conceal data in steganography. Comparative Analysis has been presented in Section 3, followed by the Conclusion.

Table I: A brief summary of various data hiding techniques in steganography.

| Cover Media | Data Hiding Techniques Proposed | Advantage(s) |
|---|---|---|
| 1. Still Image | 1. In [6] the authors implemented a scheme that uses the concept linked list of randomly embedding the data in the image and linking them together. | 1. The attacker is unable to guess the next message as the data is not hidden sequentially. Also, without the password it is not possible to access the hidden data. |
| | 2. The method proposed in [7] divided the images into equal block sizes and then uses histogram to embed the data. | 1. Rather than sending a single image containing all the hidden data, blocks of images can be sent in out of order to confuse the attacker. |

|  |  |  | 2. Since data is hidden in the histogram it is difficult to locate the data along with the increase in capacity to conceal data. |
|---|---|---|---|
|  |  | 3. Optimized Bit Plane splicing algorithm [13] is implemented where in the pixels are grouped based on their intensity and then the number of bits are to represent the hidden data are chosen. | 1. As the bits are grouped based on the intensity of the pixels, more number of darker intensity pixels can be used to represent the hidden data than just the LSB |
|  |  | 4. In [16] higher bit planes are generated by converting the pixels into different binary format using prime numbers as the weighted function. | 1. Higher bit planes can now be used to hide the data instead of just the lower bit planes. |
| 2. | Audio Signal | 1. In [17] the authors proposed two methods to use the audio signals to hide the data. In the first method, the parity of the sample is checked before replacing the LSB of the sample. In the second approach XOR-ing is carried out. | 1. It is difficult to determine the data in the audio signals because the data is not hidden directly in the sample but the parity is checked before inserting the data. |
|  |  | 2. In [19] polarity of the reverberations is added to the high frequency channels and these high frequency channels are used hide the data. | 1. As the polarity of the reverberations is used to hide the data in the high frequency signals, the stego-audio signals generated are more robust and resistant to errors during transmission. |
| 3. | IPv4 Header | 1. In [21] the authors have exploited the redundant fields like fragmentation bit and the 16-bit identification field of the IPv4 header to covertly transfer the data over the communication network. | 1. As the inherent fields of IPv4 header are utilized to transfer the data, it is extremely difficult to detect these covert channels. Hence, the sensitive data can be communicated easily by using this technique. |